# Variance for Estimation and Prediction in Linear Mixed Effects Models Under MAR

Yongqiang Tang
yongqiang_tang@yahoo.com

November 1, 2025 *

**Abstract**

In linear mixed effects models (LMM), inference on fixed effects typically relies on either the generalized least squares (GLS) variance estimator obtained by assuming that the variance parameters are known, or on the Kenward-Roger method with a small-sample bias adjustment. A simulation study shows that both approaches can underestimate the variance of the treatment effect and inflate the type I error rate in large clinical trials under missing at random (MAR) mechanisms, with greater type I error inflation as the differences in the missingness mechanisms between treatment groups increase. Counterintuitively, the estimated fixed effects and variance parameters exhibit asymptotic dependence. Ignoring this dependence may lead to invalid inference. To account for this additional source of variability, we propose a new asymptotic variance estimator and a small-sample bias-corrected estimator to quantify uncertainty in estimation and prediction in LMM under MAR. The proposed methods are theoretically justified, and their performance is demonstrated by numerical examples.

*Keywords:* Kenward–Roger variance estimator; Mixed Effects Model for Repeated Measures; Multiple Imputation; Observed information matrix; Restricted maximum likelihood

*The original version of the manuscript was submitted around November 1, 2005. The lengthy manuscript was later divided into several shorter papers for improved readability and clarity. The present paper represents the main part of that work.

# 1 Introduction

The linear mixed-effects model (LMM) is widely used as the primary method for the analysis of longitudinal continuous outcomes in clinical trials. Inference on treatment effects typically relies on either the generalized least squares (GLS) variance estimator or the Kenward-Roger (KR, Kenward and Roger (1997)) adjustment, which corrects for small-sample bias. A major challenge in these analyses is the handling of missing data. The assumption that data are missing completely at random (MCAR; Rubin (1976)) is often unrealistic. Instead, primary analyses generally assume missing at random (MAR; Rubin (1976)), whereby missing outcomes can be modeled using information from subjects with similar observed data who remained in the study (Little et al., 2012). Sensitivity analyses are then conducted to assess robustness to potential violations of the MAR assumption (Little et al., 2012).

In this paper, we investigate variance estimation in LMMs when the data are MAR. Our work is motivated by differences among variance estimators in the mixed-effects model for repeated measures (MMRM), a widely used LMM specification that imposes no specific functional form on the mean response over time and covariance structure. Empirical evidence often shows that the KR method, although more conservative than the GLS estimator, tends to yield smaller variance estimates than the multiple imputation (MI) approach in MMRM, even in large studies (Siddiqui, 2011). Using the closed-form expressions derived by Tang (2017a,b), one can show that, under a monotone missingness pattern, the expected KR variance estimator is asymptotically less than or equal to the corresponding MI variance estimator in MMRM. This asymptotic discrepancy motivates the theoretical investigation presented in this paper.

We demonstrate, through simulation and rigorous theoretical development, that the KR approach may underestimate the variability of the treatment effect and inflate the type I error rate under MAR mechanisms, potentially leading to incorrect or overly optimistic conclusions in clinical trials. To address this issue, we propose new asymptotic and bias-corrected variance estimators for estimation and prediction in LMMs. The proposed methods provide valid and reliable inference under MAR missingness, which is often the minimum assumption required for primary analyses.

The remainder of this paper is organized as follows. Section 2 develops new asymptotic and small-sample bias-corrected variance estimators for fixed effects in LMMs under the MAR assumption. The proposed estimators are compared with the existing GLS and KR variance estimators. We also suggest a method for computing the degrees of freedom (df) for inference based on $t$ or $F$ tests. In Section 3, we conduct a simulation study and analyze an antidepressant trial to evaluate the performance of different variance estimators derived under MCAR and MAR assumptions. Section 4 obtains new variance matrix formulae for the estimated fixed and random effects, enabling estimation of the mean squared error (MSE) for linear combinations of these effects. Technical proofs are provided in the supplementary material.

## 2 Variance estimators for fixed effects under MAR

### 2.1 Asymptotic variance estimators

Consider a linear mixed model (LMM) for longitudinal responses collected from $n$ subjects. Let $Y_{tot} = (Y', Y'_{mis})'$, where $Y$ denotes the observed responses, and $Y_{mis}$ denotes the missing responses. We assume that the complete data vector $Y_{tot}$ follows a multivariate normal distribution. The missing data mechanism may be either MCAR or MAR. Under MAR, the distribution of the observed data can be non-normal. For example, suppose that the complete data $(y_{i1}, y_{i2})$ for each subject follows a bivariate normal distribution, and that $y_{i2}$ is observed if and only if $y_{i1} > 0$. Among subjects with $y_{i2}$ observed, the joint distribution of $(y_{i1}, y_{i2})$ is no longer normal because the support of $y_{i1}$ is truncated.

Nevertheless, the likelihood for inference on the response model parameters based solely on the observed data remains valid under MAR and MCAR, when the parameters governing the missingness are distinct from those of the response model (Rubin, 1976). Consequently the restricted likelihood for the response model parameters based solely on the observed data remains valid under the MAR and MCAR since it can be derived from the likelihood via the Bayesian argument of Harville (1974).

For the above reasons, we continue to adopt the standard multivariate normal specification for the response model under MAR. Although the actual distribution of the observed response vector $Y$ may not be normal, the likelihood and restricted likelihood for inference on the response model parameters are identical to those obtained under the multivariate normal specification. Accordingly, we specify the following working LMM for the observed responses

$$Y = X\boldsymbol{\alpha} + Z\mathbf{u} + \mathbf{e} \tag{1}$$

where $X$ and $Z$ are known design matrices for fixed-effects $\boldsymbol{\alpha}$ and random effects $\mathbf{u}$, respectively. Assume that $X$ has full column rank $m$, and that $\mathbf{u} \sim N(\mathbf{0}, G)$ and $\mathbf{e} \sim N(\mathbf{0}, R)$ are independent. The covariance matrix for $Y$ is $\Sigma = ZGZ' + R$. For a model (e.g. MMRM) without random effects, we have $\Sigma = R$. The model can be expressed concisely as

$$Y \sim N(X\boldsymbol{\alpha}, \Sigma) \tag{2}$$

Let $\boldsymbol{\sigma} = (\sigma_1, \dots, \sigma_r)$ denote all unknown variance parameters in $\Sigma$.

The restricted maximum likelihood (REML) is often employed to reduce the bias in estimating the variance parameters in $\Sigma$. The REML log-likelihood can be derived via the Bayesian argument of Harville (1974).

$$\ell_R = \text{const} - \frac{1}{2}\log(|X'\Sigma^{-1}X|) - \frac{1}{2}\log(|\Sigma|) - \frac{1}{2}[Y - X\tilde{\boldsymbol{\alpha}}]'\Sigma^{-1}[Y - X\tilde{\boldsymbol{\alpha}}] \tag{3}$$

where $\tilde{\boldsymbol{\alpha}} = (X'\Sigma^{-1}X)^{-1}X'\Sigma^{-1}Y$ is the estimator of $\boldsymbol{\alpha}$ when $\Sigma$ is known.

Given the estimated $\hat{\Sigma}$, the REML estimator of $\boldsymbol{\alpha}$ takes the form

$$\hat{\boldsymbol{\alpha}} = (X'\hat{\Sigma}^{-1}X)^{-1}X'\hat{\Sigma}^{-1}Y \tag{4}$$

The GLS variance for $\hat{\boldsymbol{\alpha}}$ is the variance of $\tilde{\boldsymbol{\alpha}}$ derived by assuming that $\Sigma$ is known

$$\Phi_{GLS} = \Phi \tag{5}$$

where $\Phi = \text{var}(\tilde{\boldsymbol{\alpha}}) = (X'\Sigma^{-1}X)^{-1}$.

The variance of $\hat{\boldsymbol{\alpha}}$ can be obtained from the decomposition

$$\hat{\boldsymbol{\alpha}} - \boldsymbol{\alpha} = (\tilde{\boldsymbol{\alpha}} - \boldsymbol{\alpha}) + (\hat{\boldsymbol{\alpha}} - \tilde{\boldsymbol{\alpha}}) \approx (\tilde{\boldsymbol{\alpha}} - \boldsymbol{\alpha}) + \sum_{j=1}^{r} f_j(\hat{\sigma}_j - \sigma_j) \tag{6}$$

where $f_j = \frac{\partial \tilde{\boldsymbol{\alpha}}}{\partial \sigma_j} = -\Phi X'\Sigma^{-1}\frac{\partial \Sigma}{\partial \sigma_j}\Sigma^{-1}[Y - X\tilde{\boldsymbol{\alpha}}]$ (Kackar and Harville, 1984; Prasad and Rao, 1990; Harville and Jeske, 1992; Kenward and Roger, 1997; Alnosaier, 2007). When the data are MCAR, the variance of $\hat{\boldsymbol{\alpha}}$ is given by

$$\text{var}(\hat{\boldsymbol{\alpha}}) = \text{var}(\tilde{\boldsymbol{\alpha}}) + \text{var}(\hat{\boldsymbol{\alpha}} - \tilde{\boldsymbol{\alpha}}) = \Phi + \Lambda_{\boldsymbol{\alpha}} \tag{7}$$

where $W_{jk}$ is the covariance between $\hat{\sigma}_j$ and $\hat{\sigma}_k$, $P_j = -X'\Sigma^{-1}\frac{\partial \Sigma}{\partial \sigma_j}\Sigma^{-1}X$, $Q_{jk} = X'\Sigma^{-1}\frac{\partial \Sigma}{\partial \sigma_j}\Sigma^{-1}\frac{\partial \Sigma}{\partial \sigma_k}\Sigma^{-1}X$, $\text{E}[f_j] = 0$, $\text{cov}(f_j, f_k) = \Phi(Q_{jk} - P_j\Phi P_k)\Phi$, and

$$\Lambda_{\boldsymbol{\alpha}} = \text{var}(\hat{\boldsymbol{\alpha}} - \tilde{\boldsymbol{\alpha}}) = \sum_{j=1}^{r}\sum_{k=1}^{r} W_{jk}\,\text{cov}(f_j, f_k) \tag{8}$$

The above approach assumes that $\mathrm{E}[f_j] = 0$. This condition is satisfied when the marginal distribution of the observed outcomes is normal (Kackar and Harville, 1984; Prasad and Rao, 1990; Alnosaier, 2007), which is guaranteed when the data are MCAR.

As demonstrated by the numerical example in Section 3, the expectation of $f_j$ is not necessarily zero under MAR because the selection mechanism can induce departures from normality in the observed outcomes. When the data are MAR, we propose to estimate the variance of $\hat{\boldsymbol{\alpha}} - \tilde{\boldsymbol{\alpha}}$ by

$$\Gamma_\alpha = \mathrm{var}(\hat{\boldsymbol{\alpha}} - \tilde{\boldsymbol{\alpha}}) = \sum_{j=1}^{r}\sum_{k=1}^{r} f_j f_k' W_{jk} \tag{9}$$

The asymptotic variance estimator of $\hat{\boldsymbol{\alpha}}$ is given by

$$\Phi_{asy} = \Phi + \Gamma_\alpha = (X'\Sigma^{-1}X)^{-1} + \sum_{j=1}^{r}\sum_{k=1}^{r} f_j f_k' W_{ij} \tag{10}$$

Below we highlight the difference in variance estimators under MCAR and MAR

- Under MCAR, $\Lambda_{\boldsymbol{\alpha}}$ is of lower order than $\Phi$ in Equation (7). Therefore, the GLS estimator based on $\Phi$ is asymptotically valid under MCAR. Since $\Lambda_{\boldsymbol{\alpha}} = \mathrm{E}(\Gamma_\alpha)$, Equation (10) is asymptotically equivalent to Equation (7) under MCAR.

- Under MAR, $\Gamma_\alpha$ is of the same asymptotic order as $\Phi$ in Equation (10), although its magnitude may be small relative to $\Phi$. Consequently, the GLS variance estimator based on $\Phi$, which ignores $\Gamma_\alpha$, may underestimate the variance of the fixed effects when the data are MAR.

The two lemmas below establish the validity of Equation (10) for both maximum likelihood (ML) and REML estimators. We assume certain regularity conditions to ensure that both ML and REML estimators converge in probability to the true values, and are asymptotically normally distributed as the sample size $n \to \infty$.

The first lemma shows that the variance estimator in Equation (10) is asymptotically equivalent to the estimator derived under the ML framework.

**Lemma 1** *In the ML approach, the asymptotic variance of* $(\hat{\boldsymbol{\alpha}}, \hat{\boldsymbol{\sigma}})$ *is given by the inverse of the observed information matrix*

$$var\left(\begin{bmatrix}\hat{\boldsymbol{\alpha}}\\ \hat{\boldsymbol{\sigma}}\end{bmatrix}\right) = \begin{bmatrix} I_{\boldsymbol{\alpha}\boldsymbol{\alpha}} & I'_{\boldsymbol{\sigma}\boldsymbol{\alpha}} \\ I_{\boldsymbol{\sigma}\boldsymbol{\alpha}} & I_{\boldsymbol{\sigma}\boldsymbol{\sigma}}\end{bmatrix}^{-1} = \begin{bmatrix} I^{-1}_{\boldsymbol{\alpha}\boldsymbol{\alpha}} + \eta W_{\boldsymbol{\sigma}}\eta' & -\eta W_{\boldsymbol{\sigma}} \\ -W_{\boldsymbol{\sigma}}\eta' & W_{\boldsymbol{\sigma}}\end{bmatrix} \tag{11}$$

*where* $I_{\boldsymbol{\alpha\alpha}} = -\frac{\partial^2 \ell}{\partial \boldsymbol{\alpha} \partial \boldsymbol{\alpha}'} = X\Sigma^{-1}X$, $I_{\boldsymbol{\sigma\alpha}} = -\frac{\partial^2 \ell}{\partial \boldsymbol{\sigma} \partial \boldsymbol{\alpha}'}$, $I_{\boldsymbol{\sigma\sigma}} = -\frac{\partial^2 \ell}{\partial \boldsymbol{\sigma} \partial \boldsymbol{\sigma}'}$, $\eta = I^{-1}_{\boldsymbol{\alpha\alpha}} I'_{\boldsymbol{\sigma\alpha}}$ *and* $W_{\boldsymbol{\sigma}} = [I_{\boldsymbol{\sigma\sigma}} - I_{\boldsymbol{\sigma\alpha}} I^{-1}_{\boldsymbol{\alpha\alpha}} I'_{\boldsymbol{\sigma\alpha}}]^{-1}$ *is the asymptotic variance for* $\hat{\boldsymbol{\sigma}}$. *The ML approach yields an asymptotic variance for* $\hat{\boldsymbol{\alpha}}$ *that is equivalent to Equation* (10) *since* $(f_1, \dots, f_r) = -\eta + o_p(1)$.

In the ML approach, $\eta \approx -(f_1, \dots, f_r)$ does not have mean zero under MAR. Consequently, the estimated fixed effects $\hat{\boldsymbol{\alpha}}$ and variance components $\hat{\boldsymbol{\sigma}}$ cannot be assumed to be asymptotically independent when deriving their joint variance matrix from the observed information matrix. This dependence under MAR was already noted by Kenward and Molenberghs (1998) through a simple example.

Let $\begin{bmatrix} S(\boldsymbol{\alpha}) \\ S(\boldsymbol{\sigma}) \end{bmatrix} = \begin{bmatrix} n^{-1/2} \frac{\partial \ell}{\partial \boldsymbol{\alpha}} \\ n^{-1/2} \frac{\partial \ell}{\partial \boldsymbol{\sigma}} \end{bmatrix}$ be the scaled score function with respect to the ML log-likelihood, and $\mathcal{I}_{..} = \lim_{n\to\infty} \mathrm{E}[n^{-1} I_{..}]$ the per-subject expected information, evaluated under the true data-generating mechanism, where the subscripts correspond to $(\boldsymbol{\alpha}, \boldsymbol{\sigma})$. Symbols with asterisks (e.g. $\overset{*}{S}(\boldsymbol{\sigma})$ and $\overset{*}{\mathcal{I}}_{\boldsymbol{\sigma\sigma}}$) denote the corresponding functions from the REML likelihood.

Lemma 2 shows that the asymptotic variance for $(\hat{\boldsymbol{\alpha}}, \hat{\boldsymbol{\sigma}})$ in the REML approach is asymptotically equivalent to that in the ML approach.

**Lemma 2** *(a) The score function and expected information matrix evaluated at the true parameter values in the ML and REML approaches are related as follows*

$$\overset{*}{S}(\boldsymbol{\sigma}) = S(\boldsymbol{\sigma}) - \mathcal{I}_{\boldsymbol{\sigma\alpha}} \mathcal{I}^{-1}_{\boldsymbol{\alpha\alpha}} S(\boldsymbol{\alpha}) + o_p(1) \tag{12}$$

$$\overset{*}{\mathcal{I}}_{\boldsymbol{\sigma\sigma}} = \mathcal{I}_{\boldsymbol{\sigma\sigma}} - \mathcal{I}_{\boldsymbol{\sigma\alpha}} \mathcal{I}^{-1}_{\boldsymbol{\alpha\alpha}} \mathcal{I}'_{\boldsymbol{\sigma\alpha}} \tag{13}$$

*b) In the REML approach,* $\sqrt{n}(\hat{\boldsymbol{\sigma}} - \boldsymbol{\sigma}) = [\overset{*}{\mathcal{I}}_{\boldsymbol{\sigma\sigma}}]^{-1} \overset{*}{S}(\boldsymbol{\sigma}) + o_p(1)$ *is asymptotically distributed as* $N(\mathbf{0}, [\overset{*}{\mathcal{I}}_{\boldsymbol{\sigma\sigma}}]^{-1})$, *and uncorrelated with* $\sqrt{n}(\tilde{\boldsymbol{\alpha}} - \boldsymbol{\alpha}) = I^{-1}_{\boldsymbol{\alpha\alpha}} S(\boldsymbol{\alpha}) + o_p(1)$.
*c)* $\sqrt{n}(\tilde{\boldsymbol{\alpha}} - \boldsymbol{\alpha})$ *is asymptotically uncorrelated with* $\sqrt{n}(\hat{\boldsymbol{\alpha}} - \tilde{\boldsymbol{\alpha}}) \approx -\mathcal{I}^{-1}_{\boldsymbol{\alpha\alpha}} \mathcal{I}'_{\boldsymbol{\sigma\alpha}} \{\sqrt{n}(\hat{\boldsymbol{\sigma}} - \boldsymbol{\sigma})\} + o_p(1)$.
*d) The asymptotic variance of* $\sqrt{n}(\hat{\boldsymbol{\alpha}} - \boldsymbol{\alpha}) = \mathcal{I}^{-1}_{\boldsymbol{\alpha\alpha}} S(\boldsymbol{\alpha}) - \mathcal{I}^{-1}_{\boldsymbol{\alpha\alpha}} \mathcal{I}'_{\boldsymbol{\sigma\alpha}} [\overset{*}{\mathcal{I}}_{\boldsymbol{\sigma\sigma}}]^{-1} \overset{*}{S}(\boldsymbol{\sigma}) + o_p(1)$ *is*

$$var[\sqrt{n}(\hat{\boldsymbol{\alpha}} - \boldsymbol{\alpha})] = \mathcal{I}^{-1}_{\boldsymbol{\alpha\alpha}} + \mathcal{I}^{-1}_{\boldsymbol{\alpha\alpha}} \mathcal{I}'_{\boldsymbol{\sigma\alpha}} [\overset{*}{\mathcal{I}}_{\boldsymbol{\sigma\sigma}}]^{-1} \mathcal{I}_{\boldsymbol{\sigma\alpha}} \mathcal{I}^{-1}_{\boldsymbol{\alpha\alpha}} \tag{14}$$

*e)* $\hat{\alpha}$ *and* $\hat{\boldsymbol{\sigma}}$ *are asymptotically correlated*

$$cov[\sqrt{n}(\hat{\boldsymbol{\alpha}} - \boldsymbol{\alpha}), \sqrt{n}(\hat{\boldsymbol{\sigma}} - \boldsymbol{\sigma})] = -\mathcal{I}^{-1}_{\boldsymbol{\alpha\alpha}} \mathcal{I}'_{\boldsymbol{\sigma\alpha}} [\overset{*}{\mathcal{I}}_{\boldsymbol{\sigma\sigma}}]^{-1} \tag{15}$$

### 2.2 Small-sample bias correction

#### 2.2.1 Kenward-Roger approach

The KR approach (Kenward and Roger, 1997) incorporates two sources of bias correction into the GLS estimator $\Phi$ for the variance of the fixed effects. A detailed derivation of the KR variance estimator is provided by Alnosaier (2007).

Equation (8) provides the first source of bias, $\Lambda_{\boldsymbol{\alpha}}$, arising from treating the variance parameters as known (Kackar and Harville, 1984; Prasad and Rao, 1990; Kenward and Roger, 1997; Alnosaier, 2007). The second source of bias stems from estimating $\Phi$ using $\hat{\Phi}$ (Prasad and Rao, 1990; Harville and Jeske, 1992; Kenward and Roger, 1997; Alnosaier, 2007). A Taylor expansion yields the bias in $\hat{\Phi}$

$$b_{\Phi} = \mathrm{E}(\hat{\Phi} - \Phi) \approx b_{\Phi 1} - \Lambda_{\alpha} + b_{\Phi 2} = \sum_{j=1}^{r} \frac{\partial \Phi}{\partial \sigma_j} \mathrm{E}[\hat{\sigma}_j - \sigma_j] - \Lambda_{\alpha} + \frac{1}{2} \sum_{j=1}^{r} \sum_{k=1}^{r} W_{jk} \Phi \mathcal{R}_{jk} \Phi \qquad (16)$$

where $\mathcal{R}_{jk} = X' \Sigma^{-1} \frac{\partial^2 \Sigma}{\partial \sigma_j \partial \sigma_k} \Sigma^{-1} X$.

By ignoring the bias in the REML estimate $\hat{\sigma}_j$'s (i.e. $b_{\Phi 1} = 0$), Kenward and Roger (1997) obtain the bias-corrected variance of $\hat{\boldsymbol{\alpha}}$

$$\hat{\Phi}_{KR} = \hat{\Phi} + \hat{\Lambda}_{\boldsymbol{\alpha}} - \hat{b}_{\Phi} \approx \hat{\Phi} + 2\hat{\Lambda}_{\boldsymbol{\alpha}} - \hat{b}_{\Phi 2}. \qquad (17)$$

For a linear covariance structure (i.e. $\Sigma = \sum_{j=1}^{r} \sigma_j \Sigma_j + \Sigma_{r+1}$ for known $\Sigma_j$), we get $\hat{b}_{\Phi 1} = \hat{b}_{\Phi 2} = \mathbf{0}$, and

$$\hat{\Phi}_{KR} = \hat{\Phi} + 2\hat{\Lambda}_{\boldsymbol{\alpha}} \qquad (18)$$

Kenward and Roger (2009) modify their original formula by deriving the bias of $\hat{\sigma}_j$ using the Cox and Snell (1968) method, and introducing the $-b_{\Phi 1} = -\sum_{j=1}^{r} \frac{\partial \Phi}{\partial \sigma_j} \mathrm{E}[\hat{\sigma}_j - \sigma_j]$ term into Equation (17).

Note that $\hat{\Lambda}_{\boldsymbol{\alpha}}$ and $\hat{b}_{\Phi}$ are of smaller asymptotic order than $\hat{\Phi}$ in Equation (17). Consequently, the KR variance estimator is asymptotically equivalent to the GLS variance estimator, and may also underestimate the variance of the fixed effects in the presence of MAR missingness, as does the GLS estimator discussed in Section 2.1.

#### 2.2.2 Small-sample correction under MAR

The asymptotic variance estimator in Equation (10) already accounts for the variability arising from the estimation of the variance parameters. In small samples, we adjust the $\Phi$ term to correct the bias introduced by estimating $\Phi$ using $\hat{\Phi}$. Following the same strategy as the KR method, we obtain the following bias-corrected estimator

$$\Phi_{ss} = \Phi_{asy} - b_{_\Phi} \approx \Phi_{asy} + \Lambda_\alpha \tag{19}$$

for $b_{_\Phi}$ and $\Phi_{asy}$ defined respectively in Equation (16) and (10). It is generally challenging to estimate the bias of the REML estimator $\hat{\boldsymbol{\sigma}}$ under MAR except in special cases such as MMRM. We may ignore both $b_{_{\Phi 1}}$ and $b_{_{\Phi 2}}$ in Equation (19), and approximate $b_{_\Phi} \approx -\Lambda_\alpha$. We do not recommend including $b_{_{\Phi 2}}$ but ignoring $b_{_{\Phi 1}}$ for reasons given in Kenward and Roger (2009).

Lemma 3 shows that the proposed variance estimators are invariant to reparameterizations.

**Lemma 3** *Suppose $\boldsymbol{\sigma} = (\sigma_1, \dots, \sigma_r)'$ can be reparameterized through a bijective mapping to $\boldsymbol{\zeta} = (\zeta_1, \dots, \zeta_r)'$. If we recover the bias in the REML estimator $\hat{\boldsymbol{\zeta}}$ from the bias in $\hat{\boldsymbol{\sigma}}$ using*

$$E[\hat{\zeta}_j - \zeta_j] = \sum_{k=1}^{r} \frac{\partial \zeta_j}{\partial \sigma_k} E(\hat{\sigma}_k - \sigma_k) + \frac{1}{2} \sum_{k=1}^{r} \sum_{l=1}^{r} tr\left[\frac{\partial^2 \zeta_j}{\partial \sigma_k \partial \sigma_l} W_{kl}\right],$$

*the variance formulas in Equations (10) and (19) are invariant under reparameterization, regardless of whether both $b_{_{\Phi 1}}$ and $b_{_{\Phi 2}}$ are included in or omitted from Equation (19).*

### 2.3 Degrees of freedom for inference

Inference in LMMs is generally based on $t$ or $F$-tests. For testing a scalar parameter $\mathcal{L}'\boldsymbol{\alpha}$ with a known vector $\mathcal{L}$, we suggest the Satterthwaite approximation for estimating the df in the t test based on the leading term $\mathcal{L}'\Phi\mathcal{L}$ of the variance estimator $\mathcal{L}'\Phi_{asy}\mathcal{L}$ or $\mathcal{L}'\Phi_{ss}\mathcal{L}$. The resulting df coincides with that from the KR approach, but the variance estimators differ, leading to distinct $t$-statistics. For the hypothesis with a rank greater than 1, the one-moment method of Fai and Cornelius (1996) may be employed on the basis of the F test.

## 3 Numerical examples

### 3.1 Simulation

We illustrate the proposed method by analyzing 40,000 simulated randomized, placebo-controlled trials. Each trial enrolls $n = 50$ or 500 participants, equally assigned to the placebo ($g_i = 0$) or test ($g_i = 1$) treatment group. The baseline outcome $x_i$ follows a normal distribution $N(0, 100)$ truncated above at 15, so that $x_i \leq 15$. Suppose the test treatment has no effect, and the outcomes measured at two post-baseline visits 1 and 2 follow a multivariate normal distribution in both groups

$$\begin{bmatrix} y_{i1} \\ y_{i2} \end{bmatrix} \sim N\left( \begin{bmatrix} 0.6\, x_i \\ 0.48\, x_i \end{bmatrix}, \begin{bmatrix} 64 & 48 \\ 48 & 72 \end{bmatrix} \right) \tag{20}$$

Table 1: Estimated treatment effects, variance and empirical type I error by MMRM and MI. [1] sample mean and variance of estimated treatment effect at visit 2 over $40,000$ trials; [2] variance estimates (type I error %) averaged over $40,000$ simulated trials.

| | | | | MCAR [2] | | MAR [2] | | | |
|---|---|---|---|---|---|---|---|---|---|
| $n$ | $\kappa$ | $\hat{\delta}_2$ [1] | Var [1] | $\Phi_{GLS}$ | $\Phi_{KR}$ | $\Phi_{asy}$ | $\Phi_{ss}$ | $\Phi^*_{ss}$ | MI |
| 50 | 0.00 | -0.002 | 9.55 | 9.08 (5.7) | 9.42 (5.4) | 9.25 (5.5) | 9.42 (5.3) | 9.40 (5.3) | 9.59 (5.2) |
| | 0.08 | -0.003 | 10.59 | 9.35 ( **6.4**) | 9.75 (5.8) | 10.25 (5.4) | 10.45 (5.2) | 10.40 (5.2) | 10.65 (5.0) |
| | 0.16 | -0.004 | 12.71 | 9.77 (**8.3**) | 10.29 (**7.6**) | 12.30 (5.5) | 12.55 (5.2) | 12.44 (5.3) | 12.83 (5.0) |
| 500 | 0.00 | -0.010 | 0.872 | 0.868 (5.2) | 0.870 (5.1) | 0.869 (5.1) | 0.870 (5.1) | 0.870 (5.1) | 0.871 (5.1) |
| | 0.08 | -0.010 | 0.953 | 0.890 (5.9) | 0.892 (5.8) | 0.946 (5.1) | 0.948 (5.1) | 0.947 (5.1) | 0.949 (5.0) |
| | 0.16 | -0.008 | 1.113 | 0.925 (**7.5**) | 0.929 (**7.4**) | 1.109 (5.1) | 1.111 (5.1) | 1.110 (5.1) | 1.112 (5.1) |

All subjects have observed data at baseline and visit 1. Missing data at visit 2 is generated according to the following mechanism with $\kappa = 0$ (MCAR), 0.08 (MAR) and 0.16 (MAR)

$$\Pr(y_{i2} \text{ is missing}) = \begin{cases} [1 + \exp(\kappa y_{i1})]^{-1} \text{ if } g_i = 0 \\ [1 + \exp(-\kappa y_{i1})]^{-1} \text{ if } g_i = 1 \end{cases} \tag{21}$$

We analyze the data using MMRM and MI. The MMRM model uses $(y_{i1}, y_{i2})'$ as the response variable, and includes visit, baseline $\times$ visit and treatment $\times$ visit interactions as the fixed effects. An unstructured covariance matrix is used to model the within-subject errors. Following Tang (2017a), we derive closed-form analytic expressions for the analyses by reformulating the MMRM as a sequential regression. The technical details are provided elsewhere (Tang, 2026).

$$y_{ij} \sim N(\underline{\alpha}_{j0} + \underline{\alpha}_{j1} x_i + \underline{\delta}_j g_i + \sum_{k=1}^{j-1} \beta_{jk} y_{ik}, \sigma_j^2) \text{ for } j = 1, 2 \tag{22}$$

In MI, we place the prior $p(\underline{\alpha}_{j0}, \underline{\alpha}_{j1}, \underline{\delta}_j, \boldsymbol{\beta}_{jk}, \sigma_j^2) \propto (\sigma_j^2)^{-2}$ for $j = 1, 2$. The MI estimate with an infinite number of imputations is calculated using the analytic formula of Tang (2017b).

The results for the treatment effects and their variances are displayed in Table 1. Both MMRM and MI yield identical and unbiased estimates of the treatment effect $\delta_2 = \underline{\delta}_2 + \beta_{21} \underline{\delta}_1$

Table 2: Treatment effect estimates and corresponding variances at week 6 in an antidepressant trial from MMRM, cLDA and MI

| | | Variance (MCAR) | | Variance (MAR) | | | | |
|---|---|---|---|---|---|---|---|---|
| model | $\hat{\delta}_{\text{week 6}}$ | $\Phi_{GLS}$ | $\Phi_{KR}$ | $\Phi_{(asy)}$ | $\Phi_{(ss)}$ | $\Phi^*_{(ss)}$ | MI | df |
| MMRM | -2.7994 | 1.2414 | 1.2464 | 1.2432 | 1.2457 | 1.2460 | 1.2525 | 150.07 |
| cLDA | -2.7994 | 1.2056 | 1.2269 | 1.2346 | 1.2453 | 1.2456 | - | 150.76 |

at visit 2. The sample variance of $\hat{\delta}_2$ increases with $\kappa$.

Under MAR (i.e. $\kappa > 0$), both GLS and KR approaches underestimate the variance of $\hat{\delta}_2$, inflating the type I error rates ($\geq 0.074$ at the nominal 0.05 level when $\kappa = 0.016, n = 500$). The explanation is as follows. Straightforward but tedious algebra shows that

$$\hat{\delta}_2 - \tilde{\delta}_2 = (\hat{\beta}_{21} - \beta_{21})(\hat{\underline{\delta}}_1 - \hat{\underline{\delta}}_1^*) \tag{23}$$

where $\hat{\underline{\delta}}_1$ ( $\hat{\underline{\delta}}_1^*$) is estimated treatment effect at visit 1 using all randomized subjects (subjects retained at visit 2). It is easy to verify that $\text{E}(\hat{\underline{\delta}}_1) = 0$, but $\text{E}(\hat{\underline{\delta}}_1^*) \neq 0$ when $\kappa > 0$. Therefore the mean of $f_\beta = \frac{\partial \tilde{\delta}_2}{\partial \beta_{21}} = \hat{\underline{\delta}}_1 - \hat{\underline{\delta}}_1^*$ is not 0, and Equation (8) fails.

Under MAR, the proposed asymptotic method generally performs well, although it slightly underestimates the variance when $n = 50$. The two bias-corrected estimators, whether excluding ($\Phi_{ss}$) or including ($\Phi^*_{ss}$) both terms $b_{\Phi 1}$ and $b_{\Phi 2}$ (see Section 2.2.2), yield similar variance estimates in all cases and outperform the asymptotic estimator $\Phi_{asy}$, particularly when $n = 50$.

The MI variance estimator is asymptotically equivalent to the proposed estimators. MI produces slightly larger variance estimates here, since the posterior mean of $\sigma_2^2$ is unbiased in MI, but the REML estimator of $\sigma_2^2$ is biased downward in finite samples (Tang, 2017b).

Under MAR, we estimate the variance for the variance parameters based on the observed information rather than the expected information under the normality assumption. For example, when $n = 500$ and $\kappa = 0.16$, the average variance estimate for $\hat{\beta}_{21}$ over 40000 trials is 0.00294 from the observed information, and 0.00228 using the expected information (Tang, 2017a). The former is closer to the empirical variance 0.00295 for $\hat{\beta}_{21}$.

### 3.2 An antidepressant trial

We revisit the antidepressant trial analyzed by Tang (2017b). The Hamilton 17-item rating scale for depression (HAMD17) is collected at baseline and weeks 1, 2, 4, 6. The dataset consists of 172 subjects (84 on active, 88 subjects on placebo). The number of subjects with missing data is 23 in placebo and 20 in the active group. The missing data are mainly due to dropout. One intermittent missing value is imputed as 11.70432, and it is handled in the same way as Tang (2017b) prior to the formal analysis.

The data are analyzed using MMRM, constrained longitudinal data analysis (cLDA;(Lu, 2010; Liu et al., 2009)) and MI with an infinite number of imputations (Tang, 2017b). The MMRM model includes visit, baseline $\times$ visit and treatment $\times$ visit interactions as the fixed effects, and employs an unstructured covariance matrix. The cLDA model is similar to MMRM, except that baseline outcomes are included as part of the response vector in cLDA, with a common baseline mean assumed for both treatment groups.

Table 2 presents the results. In MMRM, the variance estimates for the treatment effect at Week 6 are fairly similar across methods in this example. Although the sample size in this trial is reasonably large, noticeable differences among variance estimates are observed for cLDA, with the GLS and KR variance estimates being smaller than the proposed MAR-based variance estimates. For example, the GLS variance of 1.2056 is about 3.2% lower than the bias-adjusted estimate of 1.2456 under MAR.

The conventional GLS and KR variance estimates are smaller for cLDA than for MMRM. However the proposed small-sample bias-corrected variance estimates from MMRM and cLDA become nearly identical, and agree closely with the MI variance estimate, demonstrating consistency in variance estimation in the two models.

## 4 Variance for prediction in LMM

Predicting linear combinations of fixed and random effects is relevant in various applications, including small area estimation (Prasad and Rao, 1990), animal breeding (Henderson, 1975) and clinical trials. For instance, in clinical trials where the outcome changes linearly over time within each subject, a random slope model can be used to estimate the subject-specific slope.

Let $\tau = (\boldsymbol{\alpha}', \mathbf{u}')'$ be the $(m + m_r) \times 1$ vector of all $m$ fixed and $m_r$ random effects. Note that

$$\mathbf{u}^* = \mathrm{E}(\mathbf{u}|Y) = GZ'\Sigma^{-1}(Y - X\boldsymbol{\alpha}). \tag{24}$$

If the variance parameters are known, we get

$$\tilde{\boldsymbol{\alpha}} = \Phi X'\Sigma^{-1}Y \text{ and } \tilde{\mathbf{u}} = GZ'\Sigma^{-1}(Y - X\tilde{\boldsymbol{\alpha}}) \tag{25}$$

Since $\boldsymbol{\sigma}$ is unknown in practice, we estimate $\hat{\boldsymbol{\alpha}} = \hat{\Phi} X' \hat{\Sigma}^{-1} Y$ and predict $\mathbf{u}$ by

$$\hat{\mathbf{u}} = \hat{G} Z' \hat{\Sigma}^{-1} (Y - X\hat{\boldsymbol{\alpha}}). \tag{26}$$

### 4.1 Variance of prediction errors under MCAR

This section derives the variance of $\hat{\tau} - \tau$ under MCAR. It extends the result of Harville and Jeske (1992) to a matrix framework, enabling evaluation of the MSE of linear combinations of fixed and random effects, as well as their correlations.

The variance of $\hat{\tau} - \tau$ is expressed as the total variance for $(\tilde{\tau} - \tau)$ and $(\hat{\tau} - \tilde{\tau})$. Note that $\tilde{\tau} = (\tilde{\boldsymbol{\alpha}}', \tilde{\mathbf{u}}')'$ solves the mixed model equation (Henderson, 1975)

$$F\tau = \begin{bmatrix} X'R^{-1}X & X'R^{-1}Z \\ Z'R^{-1}X & Z'R^{-1}Z + G^{-1} \end{bmatrix} \begin{bmatrix} \boldsymbol{\alpha} \\ \mathbf{u} \end{bmatrix} = \begin{bmatrix} X'R^{-1}Y \\ Z'R^{-1}Y \end{bmatrix}. \tag{27}$$

Let $\mathcal{P} = \Sigma^{-1} - \Sigma^{-1} X (X'\Sigma^{-1}X)^{-1} X'\Sigma^{-1}$. The variance of $(\tilde{\tau} - \tau)$ is given by

$$\Psi = F^{-1} = \begin{bmatrix} \Phi & -\Phi X'\Sigma^{-1} Z G \\ -G Z' \Sigma^{-1} X \Phi & G - G Z' \mathcal{P} Z G \end{bmatrix} \tag{28}$$

Under MCAR, $\mathrm{E}(\mathfrak{f}_j) = \mathbf{0}$, $\mathrm{cov}(\mathfrak{f}_j, \mathfrak{f}_k) = \mathfrak{c}_j \mathcal{P} \mathfrak{c}_k'$ and the variance of $\hat{\tau} - \tilde{\tau} \approx \sum_{j=1}^{r} \mathfrak{f}_j (\hat{\sigma}_j - \sigma_j)$ is given by (Kackar and Harville, 1984; Prasad and Rao, 1990; Harville and Jeske, 1992)

$$\Lambda_\tau = \mathrm{var}(\hat{\tau} - \tilde{\tau}) = \sum_{j=1}^{r} \sum_{k=1}^{r} \mathrm{cov}(\mathfrak{f}_j, \mathfrak{f}_k) W_{jk} \tag{29}$$

where $\mathfrak{c}_j = \begin{bmatrix} \Phi X' \Sigma^{-1} \frac{\partial \Sigma}{\partial \sigma_j} \\ -\frac{\partial G}{\partial \sigma_j} Z' + G Z' \mathcal{P} \frac{\partial \Sigma}{\partial \sigma_j} \end{bmatrix}$ is a $(m + m_r) \times N$ matrix, and $\mathfrak{f}_j = \frac{\partial \tilde{\tau}}{\partial \sigma_j} = -\mathfrak{c}_j \mathcal{P} Y$ is a $(m + m_r) \times 1$ vector.

Lemma 4 below provides the bias of $\hat{\Psi}$

**Lemma 4** *Let* $\mathfrak{R}_j = \begin{bmatrix} \frac{\partial \Sigma}{\partial \sigma_j} & \frac{\partial (ZG)}{\partial \sigma_j} \\ \frac{\partial (GZ')}{\partial \sigma_j} & \frac{\partial G}{\partial \sigma_j} \end{bmatrix}$, $\mathfrak{R}_{jk} = \begin{bmatrix} \frac{\partial^2 \Sigma}{\partial \sigma_j \partial \sigma_k} & \frac{\partial^2 (ZG)}{\partial \sigma_j \partial \sigma_k} \\ \frac{\partial^2 (GZ')}{\partial \sigma_j \partial \sigma_k} & \frac{\partial^2 G}{\partial \sigma_j \partial \sigma_k} \end{bmatrix}$ *and* $\mathfrak{b} = \begin{bmatrix} \Phi X' \Sigma^{-1} & \mathbf{0} \\ G Z' \mathcal{P} & -I \end{bmatrix}$. *A Taylor expansion yields the bias of* $\hat{\Psi}$

$$b_{\Psi} = E[\hat{\Psi} - \Psi] = b_{\Psi 1} - \Lambda_\tau + b_{\Psi 2} = \sum_{j=1}^{r} \frac{\partial \Psi}{\partial \sigma_j} E[\hat{\sigma}_j - \sigma_j] - \Lambda_\tau + \frac{1}{2} \sum_{j=1}^{r} \sum_{k=1}^{r} W_{jk} \mathfrak{R}_{jk} \tag{30}$$

*where* $\frac{\partial \Psi}{\partial \sigma_j} = \mathfrak{b} \mathfrak{R}_j \mathfrak{b}'$ *and* $\frac{\partial^2 \Psi}{\partial \sigma_j \partial \sigma_k} = \mathfrak{b} \mathfrak{R}_{jk} \mathfrak{b}' - (\mathfrak{c}_j \mathcal{P} \mathfrak{c}_k' + \mathfrak{c}_k \mathcal{P} \mathfrak{c}_j')$.

Under MCAR, we get the bias-corrected variance of $\hat{\tau}-\tau$

$$\Psi_{ss}^{(\mathrm{MCAR})}=\Psi+\Lambda_\tau-b_{_\Psi}=\Psi+2\Lambda_\tau-b_{_{\Psi1}}-b_{_{\Psi2}} \tag{31}$$

We can calculate the bias in $\hat{\sigma}_j$ using the formula of Kenward and Roger (2009), and estimate $b_{_{\Psi1}}$. For a linear covariance, $b_{_{\Psi1}}=b_{_{\Psi2}}=\mathbf{0}$. For simplicity, we may omit both $b_{_{\Psi1}}$ and $b_{_{\Psi2}}$.

### 4.2 Variance of prediction errors under MAR

**Lemma 5** *a) Under MAR, $\tilde{\tau}-\tau$ has zero mean and variance given by Equation (28).*

*b) $\mathbf{u}^*-\mathbf{u}=GZ'\Sigma^{-1}(Y-X\boldsymbol{\alpha})-\mathbf{u}$ is normally distributed and independent of $Y$ and hence $(\hat{\boldsymbol{\sigma}},\hat{\boldsymbol{\alpha}})$. This property is not affected by MAR selection governed by the observed outcome.*

*c) $\hat{\boldsymbol{\sigma}}$ is asymptotically uncorrelated with $\tilde{\boldsymbol{\alpha}}$ and hence $\tilde{\mathbf{u}}-\mathbf{u}=(\mathbf{u}^*-\mathbf{u})-GZ'\Sigma^{-1}X(\tilde{\boldsymbol{\alpha}}-\boldsymbol{\alpha})$*

*d) Ignoring the dependence between $\mathfrak{f}_j$'s and $\hat{\boldsymbol{\sigma}}-\boldsymbol{\sigma}$ yields the asymptotic variance of $\hat{\tau}-\tilde{\tau}$*

$$\Gamma_\tau=var(\hat{\tau}-\tilde{\tau})=\sum_j\sum_k\mathfrak{f}_j\mathfrak{f}_k'W_{jk} \tag{32}$$

*e) The asymptotic variance of $\hat{\boldsymbol{\tau}}-\boldsymbol{\tau}$ is $\Psi_{asy}^{(MAR)}=\Psi+\Gamma_\tau$. In finite samples, the variance is*

$$\Psi_{ss}^{(MAR)}=\Psi_{asy}^{(MAR)}-b_{_\Psi}\approx\Psi_{asy}^{(MAR)}+\Lambda_\tau \tag{33}$$

## 5 Discussion

When the data deviate from the MCAR assumption, the observed data are no longer normally distributed, and $E(Y)\neq X\boldsymbol{\alpha}$. In the presence of MAR missingness, the estimates of the fixed effects and variance parameters exhibit asymptotic dependence in LMM. Ignoring such dependence or treating the variance parameters as known leads to incorrect variance estimation. In the simulation, the conventional GLS and KR approaches underestimate the variance of the fixed-effect estimates, and inflate the type I error rates when the data are MAR. Greater type I error inflation is observed as the missingness mechanisms become increasingly different between treatment groups.

We develop asymptotic variance estimators for estimation and prediction in LMM under MAR, along with bias-corrected estimators for small-sample inference. The proposed estimators are theoretically justified and demonstrate good finite-sample performance in our numerical examples. In a companion paper, we show that the proposed variance estimators are asymptotically equivalent to both the MI (Tang, 2017b) and delta-method variance estimators (Tang, 2017a) in MMRM, thereby establishing consistency across methodologies. To ensure valid inference under MAR, we rely on the observed information matrix rather than the expected information based on the normality assumption to estimate the variance of the ML or REML estimators of the variance parameters, as demonstrated in the simulation study.

In our numerical example, the smallest sample size considered is $n = 50$. Further work, particularly large-scale simulation studies, is needed to examine the performance of the proposed methods for estimation and prediction in more complex LMM with small to moderate sample sizes.